\documentclass[aps,prd,twocolumn,twoside,superscriptaddress,nofootinbib,showpacs,showkeys,nofootinbib,notitlepage,preprintnumbers]{revtex4-1}

\usepackage[usenames,dvipsnames,svgnames,table]{xcolor} 
\usepackage{graphicx}
\usepackage{dcolumn}
\usepackage{bm}
\usepackage{hyperref}
\hypersetup{ 
    pdfnewwindow=true,      
    colorlinks=true,       
    allcolors=SeaGreen
} 
\usepackage{dsfont}
\usepackage{placeins}
\usepackage{todonotes}
\usepackage[utf8]{inputenc}
\usepackage{gensymb}
\usepackage{multirow}
\usepackage{amsmath}
	
\usepackage[capitalise]{cleveref}
\usepackage{nameref}
\usepackage{amsfonts}
\usepackage{amssymb}
\usepackage{bbm}

\usepackage{xpatch}
\makeatletter
\xpatchcmd{\@ssect@ltx}{\@xsect}{\protected@edef\@currentlabelname{#8}\@xsect}{}{}
\xpatchcmd{\@sect@ltx}{\@xsect}{\protected@edef\@currentlabelname{#8}\@xsect}{}{}
\makeatother
\usepackage{hyperref}

\usepackage{array}   
\newcolumntype{L}{>{$}l<{$}} 
\newcolumntype{R}{>{$}r<{$}}
\newcolumntype{C}{>{$}c<{$}}
\usepackage{color}
\usepackage{xspace}
\usepackage{braket}
\usepackage{units}
\usepackage{slashed}
\usepackage{multirow}
\usepackage[normalem]{ulem}
\usepackage{enumitem,amssymb}
\usepackage{xspace}

\graphicspath{ {plots/} }

\newcommand{\be}{\begin{equation}}
\newcommand{\ee}{\end{equation}}

\newcommand{\im}{\mathrm{Im}\,}
\newcommand{\re}{\mathrm{Re}\,}

\newcommand{\thline}{\hline \\[-2mm]}

\newlist{todolist}{itemize}{2}
\setlist[todolist]{label=$\square$}
\usepackage{pifont}

\usepackage{bm} 

\newcommand{\mevnospace}{\ensuremath{{\mathrm{\,Me\kern -0.1em V}}}}
\newcommand{\gevnospace}{\ensuremath{{\mathrm{\,Ge\kern -0.1em V}}}}
\newcommand{\tevnospace}{\ensuremath{{\mathrm{\,Te\kern -0.1em V}}}}

\newcommand{\mytitle}[1]{\vspace{.3cm}{\em #1.--}}

\newcommand{\sig}
{\ensuremath{\sigma/f_0(500)}\xspace}

\newcommand{\fzero}
{\ensuremath{f_0(980)}\xspace}

\newcommand{\kap}
{\ensuremath{\kappa/K_0^*(700)}\xspace}

\newcommand{\pipi}
{\ensuremath{\pi \pi \to \pi \pi}\xspace}

\newcommand{\pipikk}
{\ensuremath{\pi \pi \to K \bar K}\xspace}

\usepackage{mfirstuc} 
\newcommand{\addReviewer}[2]{
  \expandafter\newcommand\csname #1\endcsname[1]{{\bf \color{#2} \capitalisewords{#1}:\,##1}}
  \expandafter\newcommand\csname #1cor\endcsname[2]{{\color{#2} \capitalisewords{#1}:\,\st{##1}{\,\bf ##2}}}
  \expandafter\newcommand\csname #1color\endcsname{\,#2}
}

\usepackage{tikz, marginnote} 
\newcommand{\checkedby}[1]{
\ifdefined\CROSSCHECKS
  \marginnote{
    \begin{tikzpicture}
      \foreach \x [count=\xi] in {#1} {
         \node[shape=circle,inner sep=0mm,
         minimum size=2mm,
         fill=\csname \x color\endcsname] at (\xi*3mm,0) {};
       }
    \end{tikzpicture}
  }
\else
\fi
}

\usepackage{soul,color}
\definecolor{chromeyellow}{rgb}{1.0, 0.65, 0.0}
\definecolor{DodgeBlue}{rgb}{0.118, 0.565,1.000}
\definecolor{asparagus}{rgb}{0.53, 0.66, 0.42}
\definecolor{cadmiumgreen}{rgb}{0.0, 0.42, 0.24}

\addReviewer{AR}{asparagus}
\addReviewer{JR}{red}
\addReviewer{JRE}{blue}
\addReviewer{Referee}{BrickRed}

\begin{document}
\title{The $f_0(1370)$ controversy from dispersive meson-meson scattering data analyses}


\author{J.~R.~Pel\'aez}
\email{jrpelaez@fis.ucm.es}\affiliation{\ucm}
\author{A.~Rodas}
\email{arodasbi@odu.edu}\affiliation{\odu}\affiliation{\jlab}
\author{J.~Ruiz de Elvira}
\email{jacobore@ucm.es}\affiliation{\ucm}\affiliation{\bern}


\newcommand{\ucm}{Departamento de F\'isica Te\'orica and IPARCOS, 
Universidad Complutense de Madrid, 
E-28040 Madrid, Spain}
\newcommand{\bern}{Albert Einstein Center for Fundamental Physics, Institute for Theoretical Physics,
University of Bern, Sidlerstrasse 5, 3012 Bern, Switzerland}
\newcommand{\jlab}{Theory Center,
Thomas  Jefferson  National  Accelerator  Facility, 
Newport  News,  VA  23606,  USA}
\newcommand{\wm}{Department of Physics, College of William and Mary, Williamsburg, VA 23187, USA}
\newcommand{\odu}{
	Department of Physics,
	Old Dominion University,
	Norfolk, Virginia 23529, USA
}

\preprint{JLAB-THY-22-3649}
\preprint{IPARCOS-UCM-23-009}
\begin{abstract}
 We establish the existence of the long-debated $f_0(1370)$  resonance in the dispersive analyses of meson-meson scattering data. For this, we present
 a novel approach using forward dispersion relations, valid for generic inelastic resonances.  We find its pole  at $\left(1245\pm40\right)-i\,\left(300^{+30}_{-70}\right)$ MeV in $\pi\pi$ scattering. We also provide the couplings as well as further checks extrapolating partial-wave dispersion relations or with other continuation methods.
A pole at $\left(1390^{+40}_{-50}\right)-i\,\left(220^{+60}_{-40}\right)$ MeV also appears in the $\pi\pi\to K\bar K$ data analysis with partial-wave dispersion relations.
Despite settling its existence, our model-independent dispersive and analytic methods still show a lingering tension between pole parameters from the $\pi\pi$ and $K\bar K$ channels that should be attributed to data.
\end{abstract}

\maketitle

{\it Introduction.-} Quantum Chromodynamics (QCD) became the theory of strong interactions almost 50 years ago, but its low-energy regime, particularly the lightest scalar spectrum, is still under debate---see \cite{Pelaez:2015qba, Pelaez:2020gnd} and the ``Scalar Mesons below 2 GeV" note in the Review of Particle Physics \cite{pdg} (RPP).  This may be surprising since light scalars are relevant for nucleon-nucleon interactions, final states in heavy hadron decays, CP violation, the identification of the lightest glueball, and the understanding of the QCD spontaneous chiral symmetry breaking.
Moreover, a precise knowledge of this sector is not only relevant by itself and QCD, but also for the accuracy frontier of Nuclear and Particle Physics.

This debate lingers on because light scalars do not show up as sharp peaks, since some of them are very wide and overlap with others, or are distorted by nearby two-body thresholds. Indeed, their shape changes with the dynamics of the process where they appear. Hence, they must be rigorously identified from their process-independent associated poles. These appear in the complex $s$-plane of any amplitude $T(s)$ where resonances exist. Here, $s$ is the total CM-energy squared Mandelstam variable. Then, the pole mass $M$ and width $\Gamma$ are defined as $\sqrt{s_{pole}}=M-i\Gamma/2$. The familiar peak shape only appears in the real axis when the resonance is narrow and isolated from other singularities. Only then, simple Breit-Wigner (BW) approximations, K-matrices or isobar sums may be justified, but not for the lightest scalars and definitely not for the $f_0(1370)$.

Problems identifying light scalars are crudely of two types. The ``data problem'' is severe in meson-meson scattering, where scalars were first observed, since data are extracted indirectly from the virtual-pion-exchange contribution to meson-nucleon to meson-meson-nucleon scattering. Hence, the initial state is not well defined, leading to inconsistencies in the data and with fundamental principles.  This is not a problem for heavy-meson decays, generically with better statistics and less systematic uncertainty. The ``model problem'' arises when searching for poles, since analytic continuations are mathematically delicate, particularly for resonances deep in the complex plane. Unfortunately, they are often carried out with models (BW, K-matrices,``isobar" sums,...). Dispersion theory addresses both problems by discarding inconsistent data and avoiding model dependencies in data parameterizations and resonance identification.

The RPP~\cite{pdg} lists the  \sig, $f_0(980)$, $f_0(1370)$, $f_0(1500)$ and $f_0(1710)$ scalar-isoscalar resonances below 2 GeV. The longstanding controversy on the \sig existence, and the similar strange \kap, (see \cite{Pelaez:2015qba,Pelaez:2021dak}) were settled  by dispersive studies~\cite{Caprini:2005zr,GarciaMartin:2011jx,Moussallam:2011zg,Descotes-Genon:2006sdr,Pelaez:2020uiw,Pelaez:2020gnd}. The $f_0(980)$, close to $K \bar K$ threshold, has also been rigorously determined dispersively \cite{GarciaMartin:2011jx,Moussallam:2011zg}. This narrow state illustrates the process-dependence of shapes, appearing as a dip in the $\pi\pi\rightarrow\pi\pi$ cross-section but as a peak in heavy-meson decays. 
The $f_0(1500)$ and $f_0(1710)$ are well established. The former has less than 10 MeV uncertainties for its mass and width and five accurate branching fractions listed in the RPP. The $f_0(1710)$ has mass and width 
uncertainties below 20 MeV and six ``seen" decay modes.

In contrast, the $f_0(1370)$ remains controversial in the hadron community: some reviews find enough evidence to consider it well established~\cite{Bugg:2007ja} whereas other reviews \cite{Klempt:2007cp,Ochs:2013gi} or recent experiments \cite{COMPASS:2015gxz}, do not. 
Indeed, a scalar-isoscalar state between 1.2 and 1.5 GeV has been reported by several experiments~\cite{Pawlicki:1976en,Cohen:1980cq,Etkin:1982se,Akesson:1985rn,Gaspero:1992gu,Lanaro:1993km,Amsler:1994rv,Amsler:1992rx,Anisovich:1994bi,Amsler:1995bz,Amsler:1995gf,Amsler:1995bf,Abele:1996si,Abele:1996nn,Barberis:1999am,Barberis:1999ap,Bellazzini:1999sj,Aitala:2000xt,Aitala:2000xu,Abele:2001js,Abele:2001pv,Link:2003gb,Ablikim:2004wn,Garmash:2004wa,Cawlfield:2006hm,Bonvicini:2007tc,Aaij:2011fx,dArgent:2017gzv}, but with large disagreements on its parameters and decay channels. However, it was absent in the classic
$\pi\pi$ scattering analyses~\cite{Hyams:1973zf,Grayer:1974cr,Hyams:1975mc,Estabrooks:1978de}, where a resonant phase motion is not seen.
One of our main results here is that we do find such a pole using just $\pi\pi\to\pi\pi$ data and rigorous dispersive and analytic techniques. 
A pole is often found in multi-channel or multi-process data analyses, where $4\pi$ are approximated as background or quasi-two-body states
(see for instance \cite{Sarantsev:2021ein}, and 
unitarized chiral approaches in \cite{Albaladejo:2008qa, Guo:2012yt,Guo:2012ym,Ledwig:2014cla}).
In general, $f_0(1370)$ analyses suffer from some aspects of the ``model problem'' and, worse, its appearance depends strongly on the source \cite{Ropertz:2018stk,Rodas:2021tyb}.
All in all, the RPP places the $f_0(1370)$ pole within a huge range, $\left(1200-1500\right)-i\,\left(150-250\right)$ MeV, and lists its decay modes only as ``seen'', remarking the elusiveness of its two-pion coupling.
Although inappropriate for this resonance, the RPP lists its BW parameters, separating the  
``$K\bar K$ mode'' mass, always above $\sim$1.35 GeV, from the ``$\pi\pi$ mode'' mass, reaching as low as $\sim 1.2$ GeV.

Here we confirm the $f_0(1370)$ presence in $\pi\pi\to \pi\pi$ scattering data, absent in the original analyses, providing a rigorous determination of its position and coupling, using model-independent dispersive and analytic methods.
We study two-meson scattering
because it obeys the most stringent dispersive constraints.
Thus, we next explain the data dispersive constraints, then the analytic methods to reach the poles, and finally discuss results and checks. Nonessential details are given in~\nameref{app:appendix}.

{\it Dispersion relations for $\pi\pi \to \pi\pi, K\bar K$.-}  We assume the customary isospin limit. Since no bound states exist in meson-meson scattering, the fixed-$t$ amplitude $F(s,t)$,  is analytic in the first Riemann sheet of the complex-$s$ plane except for a right-hand-cut (RHC) along the real axis from $s=4m_\pi^2$ to $+\infty$. Crossing this RHC continuously leads to the ``adjacent'' sheet, where resonance poles sit. In addition, there is a left-hand-cut (LHC) from $-\infty$ to $s=-t$ due to crossed channel cuts. For forward scattering ($t=0$) and partial-waves
it extends to $s=0$. 
Using Cauchy's integral formula the amplitude in the first Riemann sheet is recast in terms of integrals over its imaginary part along the RHC and LHC.

 Customarily, the pole of a resonance with isospin $I$ and spin $J$ is obtained from $f^I_J(s)$ partial waves.  In the elastic case the adjacent sheet is the inverse of the first,
 and this is how the \sig, \fzero and \kap poles were determined dispersively \cite{Caprini:2005zr,Descotes-Genon:2006sdr,GarciaMartin:2011jx,Moussallam:2011zg,Pelaez:2020gnd}.
However, the $f_0(1370)$ lies in the inelastic region and the continuation to the adjacent sheet has to be built explicitly, for which we will use analytic techniques. To avoid model dependencies we will continue the dispersive output of our Constrained Fits to Data (CFD)~\cite{GarciaMartin:2011cn}, not the fits themselves. 

For $\pi\pi\rightarrow \bar K K$  we can use the output of partial-wave Roy-Steiner equations, recently extended to 1.47 GeV, whose corresponding CFDs were obtained in \cite{Pelaez:2018qny,Pelaez:2020gnd,Pelaez:2004vs}. 

\begin{figure}
\includegraphics[width=0.45\textwidth]{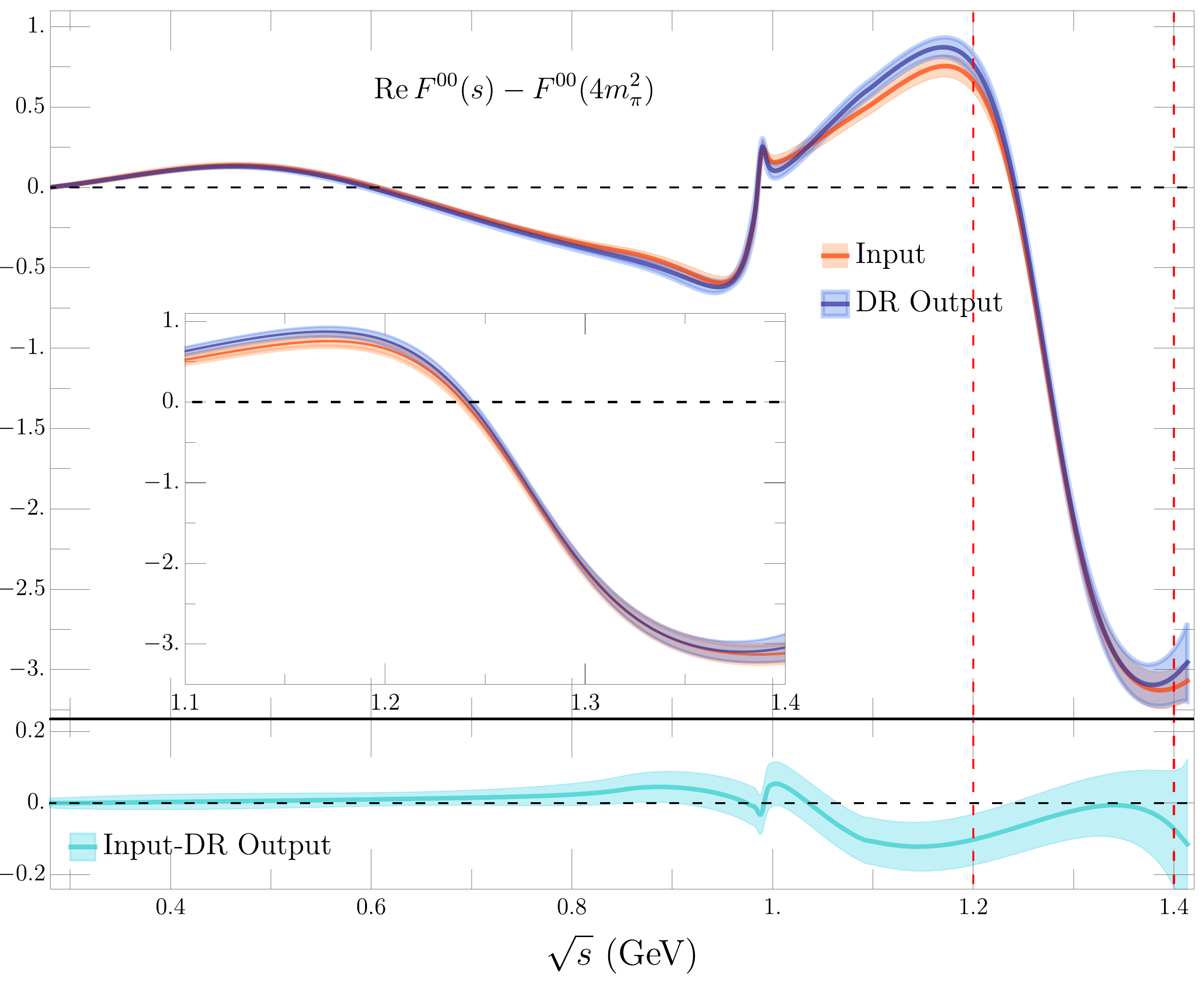} 
\caption{ $F^{00}$ Forward Dispersion Relation for the CFD \cite{GarciaMartin:2011cn}. Note the input-output agreement between 1.2 and 1.4 GeV.}\label{Fig:fdr00}
\end{figure}

However, the applicability of Roy and GKPY dispersion relations for $\pi\pi\to \pi\pi$ partial waves is limited to $\sim1.1\,$GeV. Hence, we have implemented a novel approach, by continuing the output of $\pi\pi$  Forward Dispersion Relations (FDR). These can rigorously reach any energy, and in  \cite{GarciaMartin:2011cn,Kaminski:2006qe,Pelaez:2004vs} were used to constrain partial-wave data up to 1.42 GeV.
The caveat is that FDRs alone do not determine the resonance spin.
Among the different FDRs, the most precise is that for $F^{00}\equiv(F^0+2F^2)/3$ \cite{GarciaMartin:2011cn,NavarroPerez:2015gaz},
where $F^I(s,t)$ are the $\pi\pi$ scattering amplitudes with isospin $I$. 
As input, we will use the $\pi\pi\to\pi\pi$ CFD from \cite{GarciaMartin:2011cn}, which describes data and satisfies three FDRs as well as Roy and GKPY equations \cite{Roy:1971tc,GarciaMartin:2011cn}. 
Fig.~\ref{Fig:fdr00}, shows that the once-subtracted $F^{00}$ FDR is well satisfied in the 1.2-1.4 GeV region, dominated by the $f_2(1270)$.
Note that FDRs have data input up to several tens of GeV, but above 1.42 GeV they were not used as constraints. Still, they should
remain valid within their large and growing uncertainties well above that energy.

\begin{figure}
\includegraphics[width=0.4\textwidth]{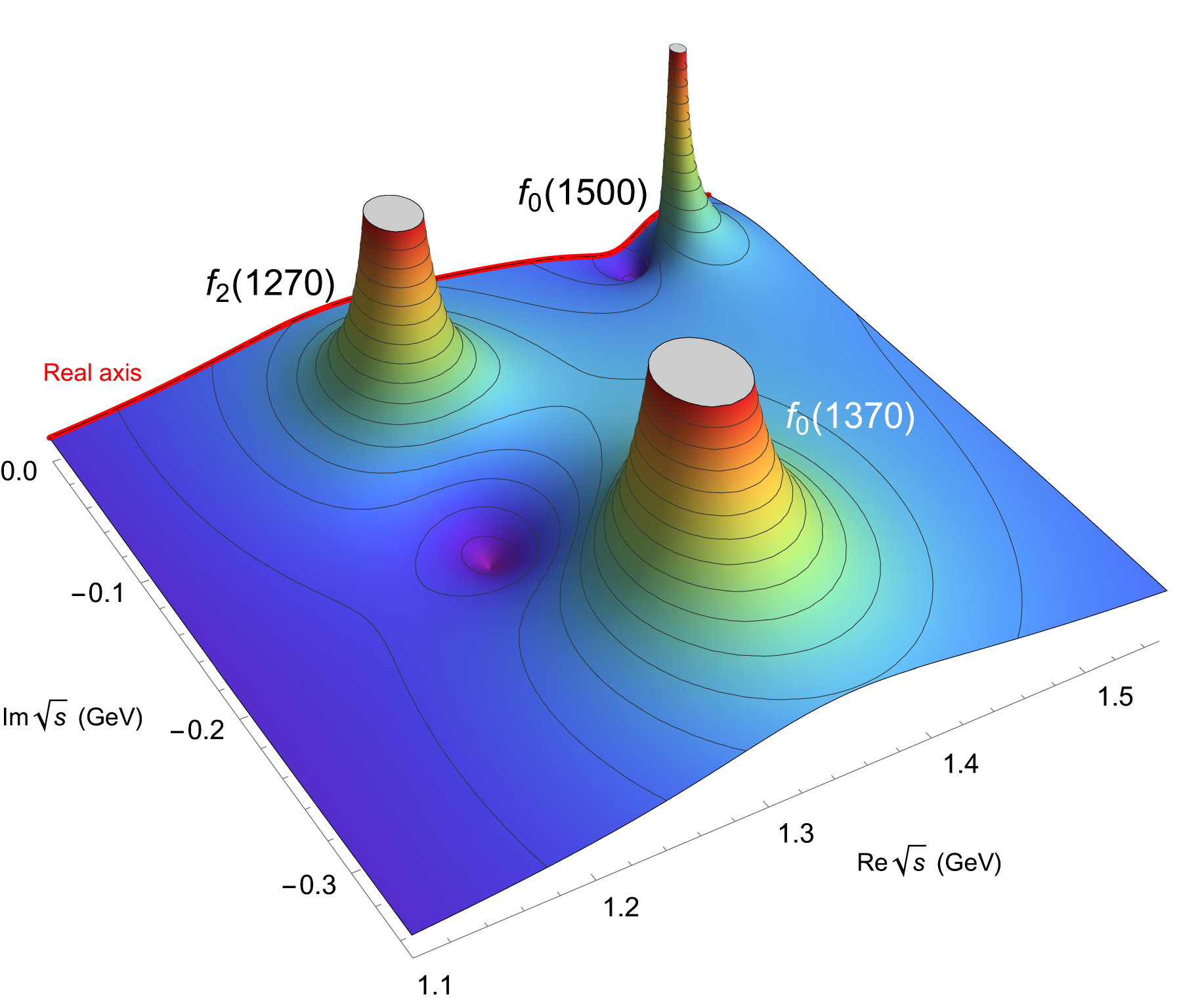} 
\caption{$\vert F^{00}(s,t=0) \vert$ obtained from 
the $F^{00}$ FDR, using as input the CFD in \cite{GarciaMartin:2011cn},  analytically continued by means of continued fractions.  Note the $f_2(1270)$ pole between the real axis and the $f_0(1370)$ pole, and the $f_0(1500)$ pole nearby.  }\label{Fig:poles0}
\end{figure}

{\it Analytic continuation methods.-}
There are several analytic continuation techniques from a real segment to the complex plane: conformal expansions~\cite{Yndurain:2007qm,Caprini:2008fc},  Laurent-Pietarinen expansions~\cite{Svarc:2013laa,Svarc:2014sqa,Svarc:2014aga}  sequences of Pad\'e approximants~\cite{Masjuan:2013jha,Masjuan:2014psa,Caprini:2016uxy,Pelaez:2016klv} or continued fractions~\cite{Schlessinger:1968,Tripolt:2016cya, Binosi:2019ecz,Binosi:2022ydc}. 
For the FDR, the $f_2(1270)$ pole lies between the real axis and the $f_0(1370)$ pole, as seen in  Fig.~\ref{Fig:poles0}. Determining the latter with Pad\'e sequences thus requires very high derivatives, making them unsuitable for the FDR method, although still valid for other checks. After trying several methods, the most stable both for $\pi\pi$ and $K \bar K$ are the continued fractions, $C_N$.  As usual, these are $N-1$ nested fractions, whose parameters are fixed by imposing $C_N(s_i)=F(s_i)$ for $N$ real values $s_i$ within the domain of interest.

{\it Results.-} 
Let us first describe the 
$\pi\pi\to\pi\pi$ $F^{00}$ FDR output continuation.
The $C_N$ are calculated from $N=7$ up to 51 equally-spaced energies in the 1.2-1.4 GeV segment, which maximizes the region where the FDR is well satisfied. Still, the $f_0(1370)$ is also found with much smaller segments, even if they lie completely below 1.3 GeV.
Fig.~\ref{Fig:poles0} shows a typical case, where $f_0(1370)$, $f_2(1270)$ and $f_0(1500)$ poles are found. Finding the latter is striking since it lies above our segment and the CFD above 1.42 GeV was not fitted to partial-wave data but to total cross-section data within a Regge formalism that describes this region ``on the average''.
  
\begin{figure}
\begin{center}
\includegraphics[width=0.48\textwidth]{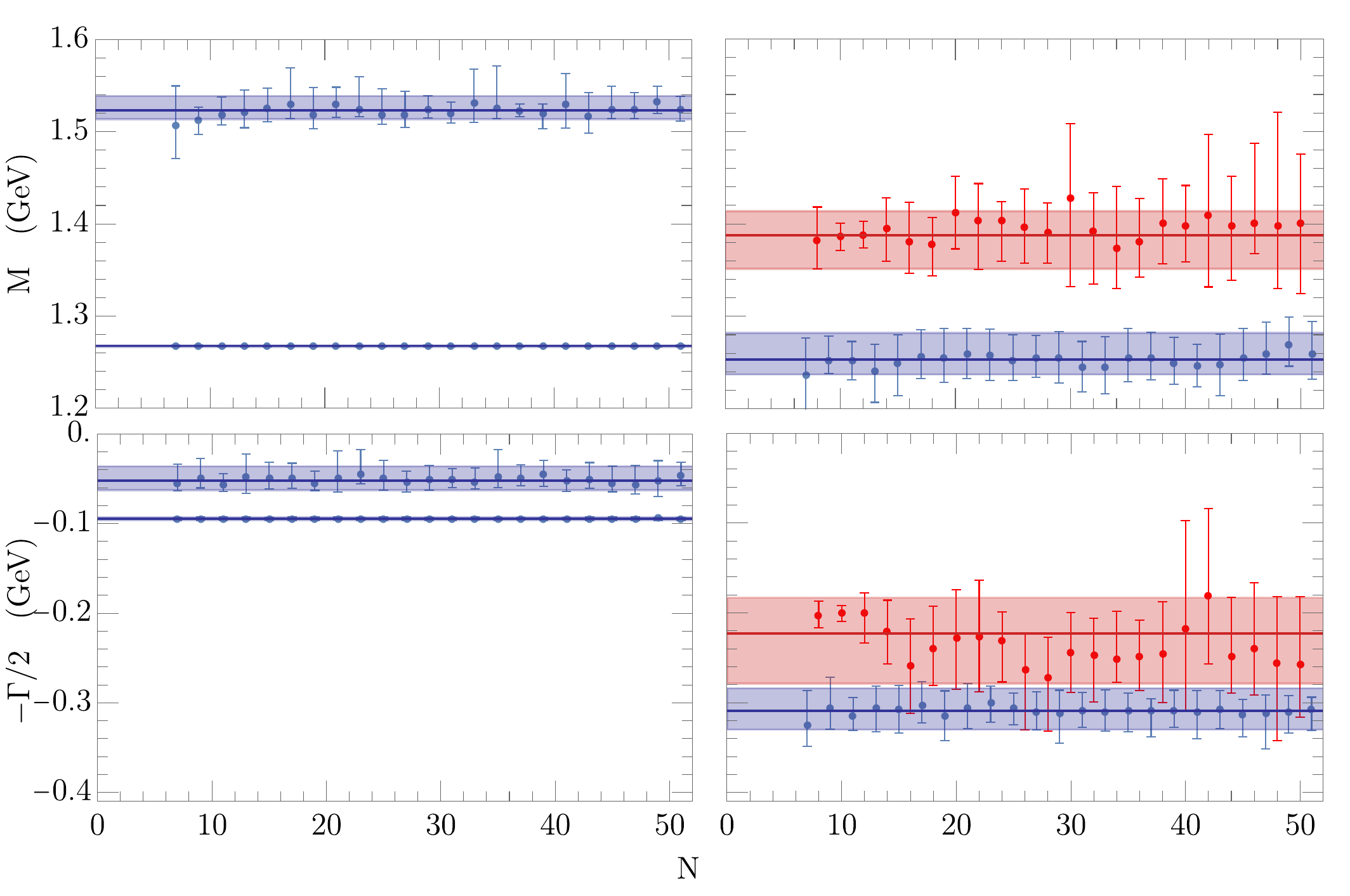}
\end{center}
\caption{  Pole masses ($M$, top) and half-widths ($\Gamma/2$, bottom)  of the  $f_2(1270)$, $f_0(1500)$ (Left) and 
$f_0(1370)$ (Right). They are obtained from the output of the $F^{00}$
$\pi\pi\to\pi\pi$ FDR (Blue) or Roy-Steiner
$\pi\pi\to K\bar K$ dispersive output (Red) analytically continued to the complex plane by a continued fraction of order $N$ (horizontal axis). CFD are used as input. Note the tension between $\pi\pi$ and $K\bar K$ $f_0(1370)$ determinations. \label{Fig:CNconv}}
\end{figure}

In Fig.~\ref{Fig:CNconv}  we show in blue the pole masses (top) and half widths (bottom) for each $N$.  Statistical errors are propagated from the CFD input \cite{GarciaMartin:2011cn,Kaminski:2006qe,Pelaez:2004vs}.
For each $N$ a systematic uncertainty is added by considering several intervals up to 25 MeV lower in either segment end. Results are very stable for the three resonances and their uncertainties are obtained from a weighted average of the values for each $N$.
 Note that the energy where the CFD tensor-isoscalar partial-wave phase reaches $\pi/2$ was fixed at 1274.5 MeV, so the $f_2(1270)$ pole appears at $1267.5-i\, 94\,$MeV, with negligible error. The $f_0(1500)$ pole is found at $1523^{+16}_{-10}-i\,\left( 52^{+16}_{-11}\right)$ MeV. As these two resonances are fairly narrow, their pole parameters are similar to their RPP BW values~\cite{pdg}.

Hence, in Tab.~\ref{tab:res} we provide the  $f_0(1370)$ pole parameters, obtained from the $\pi\pi\to\pi\pi$ FDR method. 
We assign isospin zero to this pole since a consistent but less accurate pole is also found in the $F^{I_t=1}=F^0/3+F^1/2-5F^2/6$ FDR.
FDRs alone do not fix the spin, but a consistent pole in the $\pi\pi\to \pi\pi$ scalar wave is found below with somewhat less rigorous methods.
 
\begin{table} 
\begin{tabular}{l  c c} 
Method & $\sqrt{s_{f_0(1370)}}$ (MeV)& G (GeV)\\[1.mm]
\thline
\scriptsize{FDR+CFD+$C_N$}&$\left(1253^{+29}_{-16}\right)-i\,\left( 309^{+21}_{-25}\right)$&$6.0\pm 0.3$\\[1mm]
\scriptsize{FDR+Global1+$C_N$}
&$\left(1232^{+29}_{-31}\right)-i\,\left(270^{+47}_{-32}\right)$&$4.9\pm 0.4$\\[1.mm]
\scriptsize{FDR+Global2+$C_N$}
&$\left(1227^{+27}_{-22}\right)-i\,\left(276^{+36}_{-48}\right)$&$4.9^{+0.4}_{-0.3}$\\[1.mm]
\scriptsize{FDR+Global3+$C_N$}
&$\left(1230^{+26}_{-21}\right)-i\,\left(274^{+36}_{-24}\right)$&$4.9^{+0.4}_{-0.5}$\\[1.mm]\thline
\scriptsize{$\mathbf{FDR_{\pi\pi\to \pi\pi}+C_N}$}&$\mathbf{\left(1245\pm 40\right)-i\,\left( 300^{+30}_{-70}\right)}$&$\mathbf{5.6^{+0.7}_{-1.2}}$\\[1mm]
\scriptsize{$\mathbf{RS_{\pi\pi\to K\bar K}+C_N}$}& $\mathbf{\left(1390^{+40}_{-50}\right)-i\,\left(220^{+60}_{-40}\right)}$
&$\mathbf{3.5^{+2.7}_{-1.7}}$\\[1mm]
\thline
\end{tabular}
\caption{$f_0(1370)$ pole parameters. First lines, from continued fractions on the $F^{00}$ FDR output using as input $\pi\pi\to\pi\pi$ CFD or global fits. Fifth line: final FDR+$C_N$ method result.
Last line, from partial-wave dispersion relations (RS), using as input $\pi\pi\to K\bar K$ constrained fits. 
Note that $G=\vert g_{\pi\pi}\vert$ for $\pi\pi\to\pi\pi$, whereas $G=\vert  
\sqrt{g_{\pi\pi}g_{K\bar K}}\vert$ for $\pi\pi\to K\bar K$. 
}
\label{tab:res} 
\end{table}

Concerning systematic errors, since the $\pi\pi$ CFD is a piece-wise function, we provided later three simple ``global''  analytic parameterizations~\cite{Pelaez:2019eqa}, almost identical among themselves and to the CFD up to 1.42 GeV.  Indeed, they fit Roy and GKPY equations output in the real axis and complex plane validity domains, as well as the FDRs up to 1.42 GeV. 
By construction, they contain \sig and $f_0(980)$ poles consistent with their dispersive values.
From 1.42 GeV up to 2 GeV they describe three widely different data sets, covering radically different $f_0(1500)$ scenarios. Still, Table~\ref{tab:res} shows their very similar $f_0(1370)$ poles using the same FDR+$C_N$ method.

\begin{figure}
{\resizebox{0.48\textwidth}{!}{\input{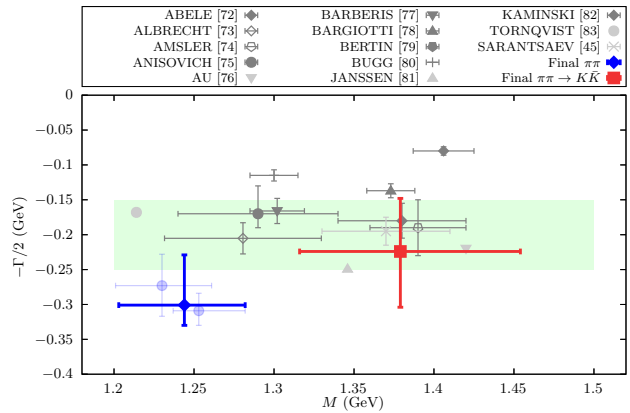}}}
\caption{ The $f_0(1370)$ poles obtained from the analytic continuation of $\pi\pi\rightarrow\pi\pi$ forward dispersion relations (blue) or $\pi\pi\rightarrow K \bar K$ partial-wave hyperbolic dispersion relations (red). For comparison, we provide the RPP $t$-matrix $f_0(1370)$ pole estimate 
(light green area) and the poles listed there \cite{CrystalBarrel:1996sqr,CrystalBarrel:2019zqh,CrystalBarrel:1995dzq,Anisovich:2009zza,Au:1986vs,WA102:1999lqn,OBELIX:2002lhi,OBELIX:1997csp,Bugg:1996ki,Janssen:1994wn,Kaminski:1998ns,Tornqvist:1995kr,Sarantsev:2021ein} (only the latest of each group, in grey). Also two poles in lighter blue, one from FDR+CFD and the other from the average of FDR+Globals. in Table~\ref{tab:res}.}
\label{Fig:f01370}
\end{figure}

For our final $\pi\pi$ FDR result, in Table~\ref{tab:res}, we first obtain a range covering all global fits, which we combine with the CFD value. In Fig.~\ref{Fig:f01370} its position is shown (in blue) in the complex $\sqrt{s}$ plane. It overlaps within uncertainties with the RPP estimate (green area), although our central half-width is $\sim50$ MeV larger.

The last row of Tab.~\ref{tab:res}
is our $f_0(1370)$ result obtained from the $\pi\pi\to K \bar K$ scalar-isoscalar partial-wave Roy-Steiner equation. Its output in the 1.04 to 1.46 GeV segment
is continued analytically by 
continued fractions. 
In Fig.~\ref{Fig:CNconv} we show, now in red, the resulting pole parameters for $N=8$ up to 50. 
Statistical uncertainties are propagated from the $\pi\pi\to K \bar K$ CFD
parameterization used as input in the integrals. For each $N$, systematic uncertainties cover the existence of two CFD solutions, three different matching points needed to describe the ``unphysical'' region between $\pi\pi$ and $K\bar K$ thresholds, and 
a variation of +30 ($-$30) MeV in the lower (upper) end of the segment. Results are stable for different $N$ and our final value is obtained by combining the (mass or width) distributions for each $N$, weighted by their uncertainties. 
Furthermore, even though the $f_2(1270)$ is not present in this wave, the uncertainties are somewhat larger than for each $\pi\pi$ row before their average. 
Nevertheless, this confirms in full rigor the $f_0(1370)$ pole existence in meson-meson scattering data and its scalar-isoscalar assignment. 

The pole position from the $\pi\pi\to K \bar K$ analysis is shown in red in Fig.~\ref{Fig:f01370}, fully consistent with the RPP estimate.
However, the central mass is around three deviations away from our $\pi\pi\to\pi\pi$ value, and the width is about one deviation away. Given the negligible model dependence of our approaches, this tension should be attributed to an inconsistency between $\pi\pi\to\pi\pi$ and $\pi\pi\to K \bar K$ data. Recall that this tension is also hinted in the RPP between the BW $\pi\pi$ and $K\bar K$ modes.

{\it Further checks}  
Roy-like partial-wave equations still hold approximately somewhat beyond their strict validity domain \cite{Ananthanarayan:2000ht,Buettiker:2003pp,Hoferichter:2015hva}. Indeed, our $f^0_0$ partial-wave Roy and GKPY dispersive output, strictly valid below 1.1 GeV, still agrees within uncertainties with the CFD input up to 1.4 GeV. The continued fraction method then yields a pole compatible with our FDR result that is listed in Tab.~\ref{tab:checks}.
being more accurate than Roy equations in that region, we provide GKPY results. Despite the unknown uncertainty due to the use of GKPY equations beyond their applicability limit, this is a remarkable consistency check, particularly for the resonance spin assignment.

\begin{table} 
\begin{tabular}{l  c c} 
Method & $\sqrt{s_{f_0(1370)}}$ (MeV)& $g_{\pi\pi}\,$(GeV)\\[1.mm]
\thline
\scriptsize{GKPY+CFD+$C_N$}
&$\left(1277_{-42}^{+49}\right)-i\,\left( 287^{+49}_{-64}\right)$&$5.6_{-2.2}^{+2.1}$\\[1.mm]
\scriptsize{GKPY+CFD+$P^N_{\;2}$}
&$\left(1285_{-36}^{+32}\right)-i\,\left( 219^{+40}_{-44}\right)$&$4.2\pm0.4$\\[1.mm]
\scriptsize{GKPY+Globals+$C_N$}
&$\left(1218_{-21}^{+26}\right)-i\,\left( 218^{+34}_{-32}\right)$&$4.1\pm1.3$\\[1.mm]
\scriptsize{GKPY+Globals+$P^N_{\;1}$}
&$\left(1220\pm27\right)-i\,\left( 213\pm26\right)$&$4.1\pm0.4$\\[1.mm]
\scriptsize{GKPY+Globals+$P^N_{\;2}$}
&$\left(1212\pm 24\right)-i\,\left( 211\pm45\right)$&$4.2\pm0.4$\\[1.mm]
\scriptsize{Globals param.+$C_N$}
&$\left(1215\!\pm\!28\right)-i\left(217\!\pm43\!\right)$& $4.2\!\pm\!0.4$\\[1.mm]
\scriptsize{Globals param.+$P^N_{\;1}$}
&$\left(1216\!\pm\!25\right)-i\left(211\!\pm43\!\right)$ & $3.9\!\pm\!0.3$\\[1.mm]
\scriptsize{Globals param.+$P^N_{\;2}$}
&$\left(1220\!\pm\!25\right)-i\left(207\!\pm40\!\right)$ & $3.8\!\pm\!0.4$\\[1.mm]
\scriptsize{Globals param.}
&$\left(1215\!\pm\!29\right)-i\left(217\!\pm45\!\right)$ & $4.2\!\pm\!0.4$\\[1.mm]
\thline
\end{tabular}
\caption{Approximated methods yield poles compatible with the rigorous $\pi\pi$ results in Table~\ref{tab:res}. We compare results using CFD as input and the average of using the global parameterization as inputs
as well as different continuation methods: Pad\'e sequences ($P_M^N$), continued fractions ($C_N$) or directly from the global parameterization. GKPY equations are extrapolated beyond their strict validity range. 
}
\label{tab:checks} 
\end{table}

Moreover, since in the $f^0_0$ partial wave there is no $f_2(1270)$ pole hindering the $f_0(1370)$ determination, Pad\'e sequences provide a check with a different continuation method.
Recall that a Pad\'e approximant of $f(s)$ is
$P^N_{M}(s,s_0)=Q_N(s,s_0)/R_M(s,s_0)$,
with $Q_N$ and $R_M$ polynomials of $N^{th}$ and $M^{th}$ degree, respectively, matching 
the $f(s)$ Taylor series to order $N+M+1$. Namely,
$P^N_{M}(s,s_0)=f(s)+{\cal O}\left((s-s_0)^{N+M+1}\right)$.
The polynomial coefficients are related to the $f(s)$ derivatives of different orders. It has been proved~\cite{Ballore:1902} that if $f(s)$ is regular inside a domain $D$, except for poles at $s_{p_i}$, of total multiplicity $M$, the sequence $P^N_M(s)$ converges uniformly to $f(s)$ in any compact subset of $D$, excluding the $s_{p_i}$. 
The  Pad\'e sequence choice depends on the partial-wave analytic structure. In our case, at least it must have a pole for the resonance, although we also considered sequences with more poles. We follow previous works \cite{Pelaez:2016klv}, now using as input the GKPY output. We have propagated the data uncertainties and added systematic errors from the sequence truncation and $s_0$ choice. As seen in Tab.~\ref{tab:checks}, the pole from the GKPY output continued with the $P^N_2$ sequences is consistent with that from continued fractions. Similar consistency is found for other Pad\'e sequences. 
 
In addition, we have checked the consistency and accuracy of the dispersive plus continuation methods versus the global parameterizations since, being analytic, they can be directly extended to the complex plane without continuation methods. Although not built for that, these parameterizations possess an $f_0(1370)$ pole, identical up to a few MeV, even if they differ widely among themselves above 1.42 GeV. For illustration, in Tab.~\ref{tab:checks} we also list the average pole of the results obtained using the ``Global" parameterizations.
 This is a simple but parameterization-dependent extraction, remarkably close to our dispersive result, although somewhat narrower. In Tab.~\ref{tab:checks}  we also list poles obtained from its GKPY dispersive output continued to the complex plane, either with continued fractions or different Pad\'e sequences.
All of them come very close to the direct result, although with larger uncertainties, which also happens for the other global parameterizations.
Interestingly, when using Pad\'e sequences, there is also a $f_0(1500)$ pole,  with large uncertainties. Note that the three global parameterizations cover generously the $f_0(1500)$ scenarios without a significant $f_0(1370)$ change. 
Finally, the dispersive integral provides rigorously the FDR in the upper half complex plane. Our continued fraction method agrees, within less than half its uncertainty, with this dispersive result in the region $1.2\leq\re \sqrt{s}\leq1.5\,$GeV and
$\im \sqrt{s}\leq0.5\,$GeV, even before adding systematic uncertainties.
All these checks confirm the robustness of our approach.

{\it Summary.-} We have presented a method, combining analytic continuation techniques with forward dispersion relations, to find poles and determine accurately their parameters avoiding model dependencies, even in the inelastic regime. This provides rigorous dispersive results  beyond the validity range of conventional partial-wave equations.
Applied to  $\pi\pi$ scattering, this method 
reproduces the $f_2(1270)$ resonance and settles the long debate about the existence of an $f_0(1370)$ pole 
in the $\pi\pi\to\pi\pi$ amplitude, absent in the original experimental analyses. It is found at
$\left(1245\pm40\right)-i\,\left(300^{+30}_{-70}\right)$ MeV. The method also provides its elusive $\pi\pi$ coupling $\vert g_{\pi\pi}\vert=5.6^{+0.7}_{-1.2}\,$GeV.
Remarkably, it also displays an $f_0(1500)$ pole, although no partial-wave data are used in that region, just total cross-section data. 
Consistent results are obtained in the extrapolation of usual Roy-like dispersion relations. 
Finally, a $f_0(1370)$ pole at $\left(1390^{+40}_{-50}\right)-i\,\left(220^{+60}_{-40}\right)$ MeV is also found in the continuation of hyperbolic partial-wave dispersion relations for $\pi\pi\to K\bar K$ scattering, showing approximately a three-sigma tension in the mass that can only be attributed to data.

The method presented here can be easily applied to other processes to avoid the pervasive model dependence in hadron spectroscopy and opens the possibility of using total cross-section data
avoiding partial-wave analyses.


\begin{acknowledgments} 
  This project has received funding from the Spanish Ministerio de Ciencia e Innovación grant PID2019-106080GB-C21 and the European Union’s Horizon 2020 research and innovation program under grant agreement No 824093 (STRONG2020). AR acknowledges the financial support of the U.S. Department of Energy contract DE-SC0018416 at the College of William \& Mary, and contract DE-AC05-06OR23177, under which Jefferson Science Associates, LLC, manages and operates Jefferson Lab. J. R. E. acknowledges financial support from the Swiss National Science Foundation under Project No. PZ00P2174228 and from the Ramón y Cajal program (RYC2019-027605-I) of the Spanish MINECO.
\end{acknowledgments}

\appendix
\section*{\bf Appendix}\label{app:appendix}

\subsection{Continued fractions and Pad\'e approximants}\label{app:cont}

Continued fractions have been used in many different areas of physics with great success. In particular, the use of continued fractions for the analytic continuation
of scattering amplitudes dates back to the late 1960's \cite{Schlessinger:1968}, where it was specifically applied to the two-body scattering case, even with two channels.
In the recent past they have also been found useful in other hadron physics applications \cite{Tripolt:2016cya, Binosi:2019ecz,Binosi:2022ydc}. 

In our case, we want to obtain an analytic continuation to the complex $s$-plane from the information on the amplitude within a segment in the real axis. Hence, given $N$ real energy-squared values $s_i$, the continued fraction $C_N$ for our FDR amplitude $F$ satisfies $C_N(s_i)=F(s_i)$ and is defined as
\begin{equation}
C_{N}(s)=F\left(s_{1}\right)\Big/\Big(1+\frac{a_{1}\left(s-s_{1}\right)}{1+\frac{a_{2}\left(s-s_{2}\right)}{\ddots a_{N-1}\left(s-s_{N-1}\right)}}\Big). 
\end{equation}
For example, $C_1(s)=F(s_1)$, a constant; $C_2(s)=F(s_1)/(1+a_1(s-s_1))$, etc...
Note that depending on whether $N$ is even or odd, $C_N$ tends to zero or to a constant at $s\to \infty$, respectively. 

The $a_i$ parameters can be obtained recursively as
\begin{align}
a_{1}&=\big(F\left(s_{1}\right)/F\left(s_{2}\right)-1\big)\big/\big(s_{2}-s_{1}\big),  \nonumber \\
a_i  &= \frac{1}{s_i-s_{i+1}}\Big(1+\frac{a_{i-1}(s_{i+1}-s_{i-1})}{1+\frac{a_{i-2}(s_{i+1}-s_{i-2})}
{\ddots \frac{a_{1}(s_{i+1}-s_{1})}{1-F\left(s_{1}\right)/F\left(s_{i+1}\right)}}}\Big),
\end{align}
and therefore $a_i$ contains information from  $i+1$ points, namely, $s_1$ up to $s_{i+1}$.

The other continuation method we have used is that of sequences of Pad\'e approximants~\cite{Masjuan:2013jha,Masjuan:2014psa,Caprini:2016uxy,Pelaez:2016klv}. These are rational-fraction approximations to a function $f$ at a given value $s$, defined as
\begin{equation}
    P^N_{M}(s,s_0)=\frac{Q_N(s,s_0)}{R_M(s,s_0)},
    \label{eq:Padedef}
\end{equation}
with $Q_N$ and $R_M$ polynomials of degree $N$ and $M$, respectively, satisfying
\begin{equation}
     P^N_{M}(s,s_0)=f(s)+{\cal O}\left((s-s_0)^{N+M+1}\right), 
     \label{eq:PadeTaylor}
\end{equation}
i.e., they match the Taylor series expansion (around $s_0$) of $f$ at order $N+M+1$, hence ensuring that their coefficients are unique. 
As explained in the main text, Montessus de Ballore's theorem~\cite{Ballore:1902},  states that if $f(s)$ is regular inside a domain $D$, except for poles at $s_{p_i}$, of total multiplicity $M$, then the sequence $P^N_M(s)$,
with $N\rightarrow \infty$, converges uniformly to $f(s)$ in any compact subset of $D$, excluding the $s_{p_i}$. 
Thus, this sequence provides the analytic continuation to the adjacent Riemman sheet where we can look for poles in a model-independent way~\cite{Masjuan:2013jha,Masjuan:2014psa,Caprini:2016uxy,Pelaez:2016klv}. 

In practice, the choice of $M$ 
depends on the analytic structure of each partial wave in the domain of interest in the complex plane, i.e., on its singularities, branch points and proximal resonance poles, in the lower half of the adjacent Riemann sheet.

Finally, let us remark  that a continued fraction can be understood as a Pad\`e approximant of order $(N/2-1,N/2)$ or $((N-1)/2,(N-1)/2)$,
with $N/2$ or $(N-1)/2$ poles in the complex plane, depending on whether $N$ is even or odd, respectively.
The fact that both the degrees of the numerator and denominator increase uniformly with the number of points prevents us from invoking for continued fractions the same theorems that prove the uniform convergence to $F(s)$ of the Pad\'e approximants.

\subsection{Details of numerical results}\label{app:results} 

Here, we will detail the choice of parameters of our continuation methods, the search for poles and the calculation of uncertainties.

\mytitle{ Continued fractions} 
Let us first discuss the length of the interval and the number of inner points $N$ to be interpolated. Later on, we will explain how the systematic uncertainties due to these choices are combined with the statistical uncertainties from the CFD or global input parameterizations.

Ideally, the length for the interpolation interval should be the largest segment where the resonance of interest dominates the amplitude. Unfortunately, the $f_0(1370)$ does not have a clear peak, and it does not dominate the amplitude. Hence, the best we can do is to choose a large area in the region of interest, where the dispersion relations are best satisfied. Thus, looking at Fig.~1 in the main text, we see that the $\pi\pi\to \pi\pi$ CFD  satisfies the $F^{00}$ Forward Dispersion Relation within uncertainties
in the $\sim$1.2 to 1.4 GeV region, which we will consider our reference interval. However, we will add a systematic uncertainty by 
considering several intervals up to 25 MeV lower in either segment end. We cannot take the interval higher because the FDRs were imposed on the CFD only up to 1.42 GeV and we do not want to get too close to the end region, which is naturally less stable.
All these considerations apply the same to the $\pi\pi$ global fits.
In contrast, for $\pi\pi\to K \bar K$ 
we take the interval 1.04 to 1.46 GeV as our central choice, since 
Roy-Steiner equations
are well satisfied there (see \cite{Pelaez:2020gnd}), and consider a variation of +30 (-30) MeV in the lower (upper) end of the segment to estimate a systematic uncertainty.

Next, we have to discuss the number $N$ of  $\sqrt{s_i}$ points, to be interpolated on each segment. 
First of all, for  $\pi\pi\to\pi\pi$,
we are dealing with the $F^{00}$ FDR, which is not expected to go to zero at $s\to\infty$ \footnote{The $F^0_0$ amplitude is dominated by the Pomeron at high energies, which definitely does not tend to zero. Of course, it is now known that the Pomeron contribution grows logarithmically, but the original Gribov-Pomeranchuk~\cite{Gribov:1962fw} proposal tends to a constant and is a good approximation up to roughly 15 GeV (see \cite{Pelaez:2003ky}), well above our region of interest. A pure logarithmic singularity and its growth can only be mimicked with an infinite series, but of course, our $N$ is finite}.
Thus, in this case, we consider odd values of $N$.
Nevertheless, we have found that using an even $N$ does not change much the central value but yields a significantly larger uncertainty.
In contrast, partial waves with different initial and final states, as the $\pi\pi\to K \bar K$, generically tend to zero at $s\to\infty$.
Thus, we now take even values of $N$, although considering odd values yields similar results, with slightly larger uncertainties.

What range of $N$ values should we consider?
On the one hand, a too-small $N$ may not provide enough information on the amplitude. On the other hand, a too-large $N$, with its many parameters, may give rise to numerical problems or even artifacts
due to a loss of accuracy when calculating iteratively the coefficients of the continued fraction \cite{Tripolt:2016cya}.
 Taking into account that the $f_2(1270)$ and $f_0(1500)$ resonances are well established and we need enough flexibility to accommodate the putative $f_0(1370)$ pole,  we need at least three possible poles and therefore $N\ge6$. This is why in Fig.~3 in the main text we are showing the resulting values of the mass and width of these resonances 
 varying $N$ from 7 to 51 for the $\pi\pi\to\pi\pi$
 FDR extraction (in blue). For $\pi\pi\to K\bar K$ scattering only two resonances are present and hence, one could in principle start from $N\ge 4$.  Nevertheless, we only find stable results in the range 8 to 50, (in red in Fig.~3 in the main text).
 
 Let us also recall that the number of poles in $C_N$ grows with $N$, but only the ones we show appear consistently in the region of interest,
near the 1.2 to 1.4 segment in the real axis.
In particular, in Fig.~3 in the main text, we see that already for $N=7$ we do find the three poles in that region. 
This would mimic a very simple model with three poles, but by considering larger numbers of parameters---and we are considering $N$ up to 50 or 51--- we can accommodate any possible model, rendering, in practice, our whole approach model-independent. 

All in all, for $\pi\pi$ scattering we look for the poles that appear when considering 23 odd values of $N$ from 7 to 51, 26 samples of segments with different ends, and 119 sets of CFD input parameters, including their central values and the variation due to 
changing each one of the 59 CFD parameters within its uncertainty. This is a sample of 71162 configurations.
Of course, having so many parameters and numerous nested denominators one can generate artifacts for some of these samples. Thus we only consider as model-independent features that are robust and stable under the variation of $N$, segments, etc. Cases when artifacts appear for just one choice of parameters, or interval, or are unstable under the change of $N$ are discarded. Out of those 71162 configurations, our numerical algorithm finds the $f_2(1270)$ pole, the one clearly visible to the naked eye, in all cases but one, the $f_0(1370)$ in $\sim 97\%$ of the samples and the $f_0(1500)$ in $\sim 96\%$ of them. This does not mean that the $f_0(1370)$ pole is not present in $3\%$ of the sample, but just that our automatized algorithm fails to detect them. We have checked by visual inspection that in many of those missing cases the pole is still there. Only a few samples have real artifacts.
These numbers are similar for the global fits.

We have now estimated uncertainties in three different ways.
The simplest and most naive approach would be to average all the valid samples. If we do this for the $f_0(1370)$ we find:
\begin{eqnarray}
\sqrt{s_{f_0(1370)}}&=&\left(1.255\pm19\right) -i\,\left(0.309\pm 18\right)\, {\rm GeV},\nonumber\\ g_{\pi\pi}&=&6.0\pm0.7.
\end{eqnarray} 

However, we think that not all the samples should be weighted the same, since the uncertainties coming from the CFD parameters have a statistical nature but those due to the choice of $N$ and segment are systematic.
In addition, for each choice of $N$ and segment, the statistical uncertainty differs and it could even be asymmetric. So, we have used a slightly more elaborate procedure separating the systematic from the statistical errors, using the latter to weigh the sample. The result is very similar, differing in the few MeVs both for the central values and the uncertainties. However, we prefer this more sophisticated procedure, which we have used to give our final results in the main text and that we describe next.

Let us then denote by $X$ any of the quantities we want to determine, i.e., the pole mass, width, or residue of any of the resonances we are interested in.
Then, for a given number $N$ of interpolation points, and a given interval, labeled $k$, we vary the parameters of the parameterization, we look for the quantity in question,
and obtain its central value $X_{N,k}$ and statistical uncertainty $\Delta X_{N,k}$. 
We repeat for different intervals 
obtaining different pole values and uncertainties. The difference between these values is not only due to statistics but also to the systematic effect of changing the interval. Our central value for this $N$ is then the weighted mean of all these determinations, i.e.,
\begin{align}\label{eq;ponderateN}
X_{N}=\frac{\sum\limits_{k}{X_{N,k} \,w_{N,k}}}{\sum\limits_{k}w_{N,k}},\quad
w_{N,k}\equiv\frac{1}{\left(\Delta X_{N,k}\right)^2}.
\end{align}

The systematic error associated with considering different intervals is then estimated through the weighted standard deviation 
\begin{equation}
\left(\Delta X_N^{\text{sys}}\right)^2=\frac{\sum\limits_{k}{\left(X_{N,k}-X_N\right)^2w_{N,k}}}{\sum\limits_{k}w_{N,k}-\sum\limits_{k}w_{N,k}^2\big{/}\sum\limits_{k}w_{N,k}}.
\end{equation}
Finally, the total error for a given $N$ is defined as the sum of this systematic error and the minimum of the statistical errors 

\begin{equation}
\Delta X_N=\Delta X_N^{\text{sys}}+\min_k{\Delta X_{N,k}}.
\end{equation}
Since the statistical errors are asymmetric, in practice we obtain two different central values, resulting from considering either the upward or downward uncertainty for the weights $w_{N,k}$. In Fig. 3
in the main text, the final $X_N$ are indeed their average, and
we add to the uncertainty half of the difference between
both values. These are the central values and vertical error bars shown in Fig.~3 in the main text for each resonance pole mass and width, for different values of $N$. There we see that the pole parameters of the $f_2(1270)$, $f_0(1500)$ and $f_0(1370)$ are very stable under changes of $N$, both in their central values and the size of their uncertainties. Moreover, by increasing $N$, we do not find more than these three robust poles in the region under study.

The final values and errors collected in Tables~1 and~2 in the main text are obtained in a similar way. Namely, the central value is defined from the weighted average of the determinations at different $N$ values
\begin{align}
X=\frac{\sum\limits_{N}{X_{N} \,w_{N}}}{\sum\limits_{N}w_{N}},\quad
w_{N}\equiv\frac{1}{\left(\Delta X_{N}\right)^2}.
\end{align}
Here we propagate separate $X_N$ for
upward and downward uncertainties and their average provides the central line of each quantity in Fig.~3 in the main text.
The spread between the different $X_{N}$ values for each of these two sets is again associated with a systematic error defined as
\begin{equation}
\Delta X^{\text{sys}}=\frac{\sum\limits_{N}{\left(X_{N}-X\right)^2w_{N}}}{\sum\limits_{N}w_{N}}.
\end{equation}

Finally, the total uncertainty is estimated by adding linearly both the previous systematic error and the minimum statistical uncertainty for every $N$
\begin{equation}
\Delta X=\Delta X^{\text{sys}}+\min_N{\Delta X_{N}}.
\end{equation}
Once again, due to the asymmetry between upward and
downward uncertainties, our final estimate is an average
of the two different $X$ values, and we add half their difference as an extra systematic error. 
This corresponds to the width of the bands in Fig.~3  and our results in  Tables~1 and~2 in the main text. For \pipikk, out of the two CFD sets presented in~[2], we take the CFD results obtained from the favored ``Combined'' solution, using three different matching points $\sqrt{t_m}=1.1,\,1.2,\,1.3$ GeV for the dispersion relation, to generate different input sets for our interpolators, which are then averaged according to the method described above. Moreover, we add the difference of the central values with respect to the second solution as an extra systematic uncertainty.

We also implemented a third alternative for estimating uncertainties. We start again from the value of one of the parameters of interest $X$ obtained from a particular segment $k$ and number of pints $N$ in the continued fraction, $X_{N,k}$, and its associated error $\Delta X_{N,k}$. If this error is symmetric one assumes a normal distribution to describe the problem, namely:
\begin{equation}
f_{nd}(x)=\frac{1}{\sigma \sqrt{2 \pi}} e^{-\frac{1}{2}\left(\frac{x-\mu}{\sigma}\right)^{2}},
\end{equation}
with a cumulative distribution function ($\mathcal{CDF}$) given by
\begin{equation}
\mathcal{CDF}_{nd}(x)=\int_{-\infty}^{x} f_{nd}(t) dt.
\end{equation}
However, if the error is asymmetric we assume it can be associated with a skew-normal distribution 
\begin{equation}
f_{snd}(x)=2 f_{nd}(x) \mathcal{CDF}_{nd}(\alpha x),
\end{equation}
with parameter $\alpha$.

In this spirit, one could describe the result of every interpolator $(N,k)$ with such a distribution. The final step would be to add the different $\mathcal{CDF}$ according to a normalizing weight, as for example 
\begin{align}
\mathcal{CDF}_{N}(x)&=\frac{\sum\limits_{k}{\mathcal{CDF}_{N,k}(x) \,w_{N,k}}}{\sum\limits_{k}w_{N,k}},\nonumber \\
w_{N,k}&\equiv\frac{1}{\left(\Delta X_{N,k}\right)^2},
\end{align}
where the error can be the upward or downward uncertainty or an average of the two. The method is then iterated by averaging over $N$. Finally, once the final $\mathcal{CDF}$ is computed one should calculate the median and confidence intervals to produce the final errors. Despite the fact that this method looks different from the previous one, their results are perfectly compatible, thus confirming the robustness of our error estimates.

The difference of any of these two ``weighted" methods with the naive method of averaging the whole sample without any weighting is therefore in the few-MeV range both for the central values and the uncertainties, which we can also make asymmetric. We have preferred to consider the weighted methods for our final results. Furthermore, the difference basically disappears when we round our final central values and uncertainties up to the tens.

\begin{figure}[h]
{\centering
\includegraphics[width=0.45\textwidth]{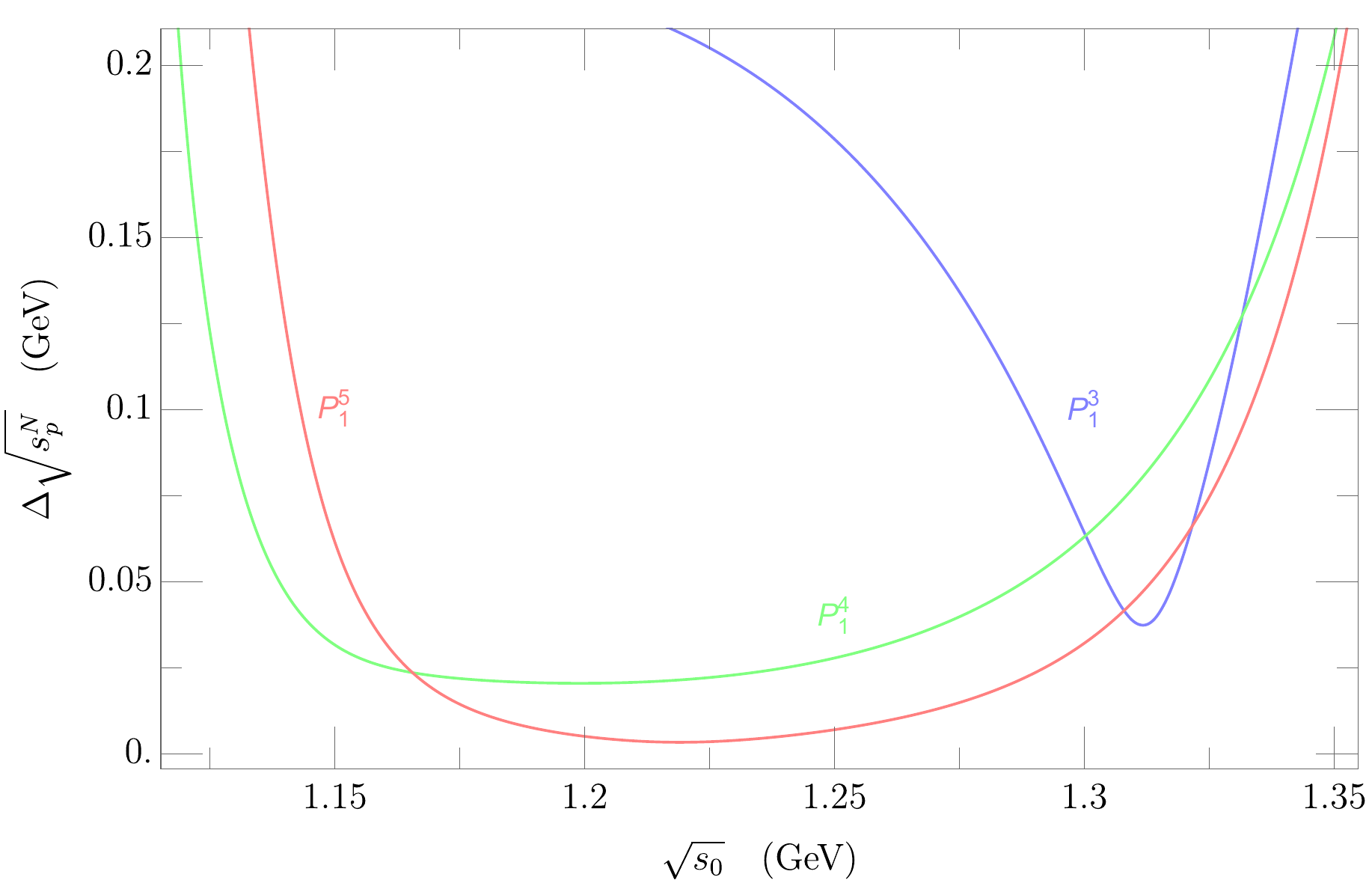} \includegraphics[width=0.45\textwidth]{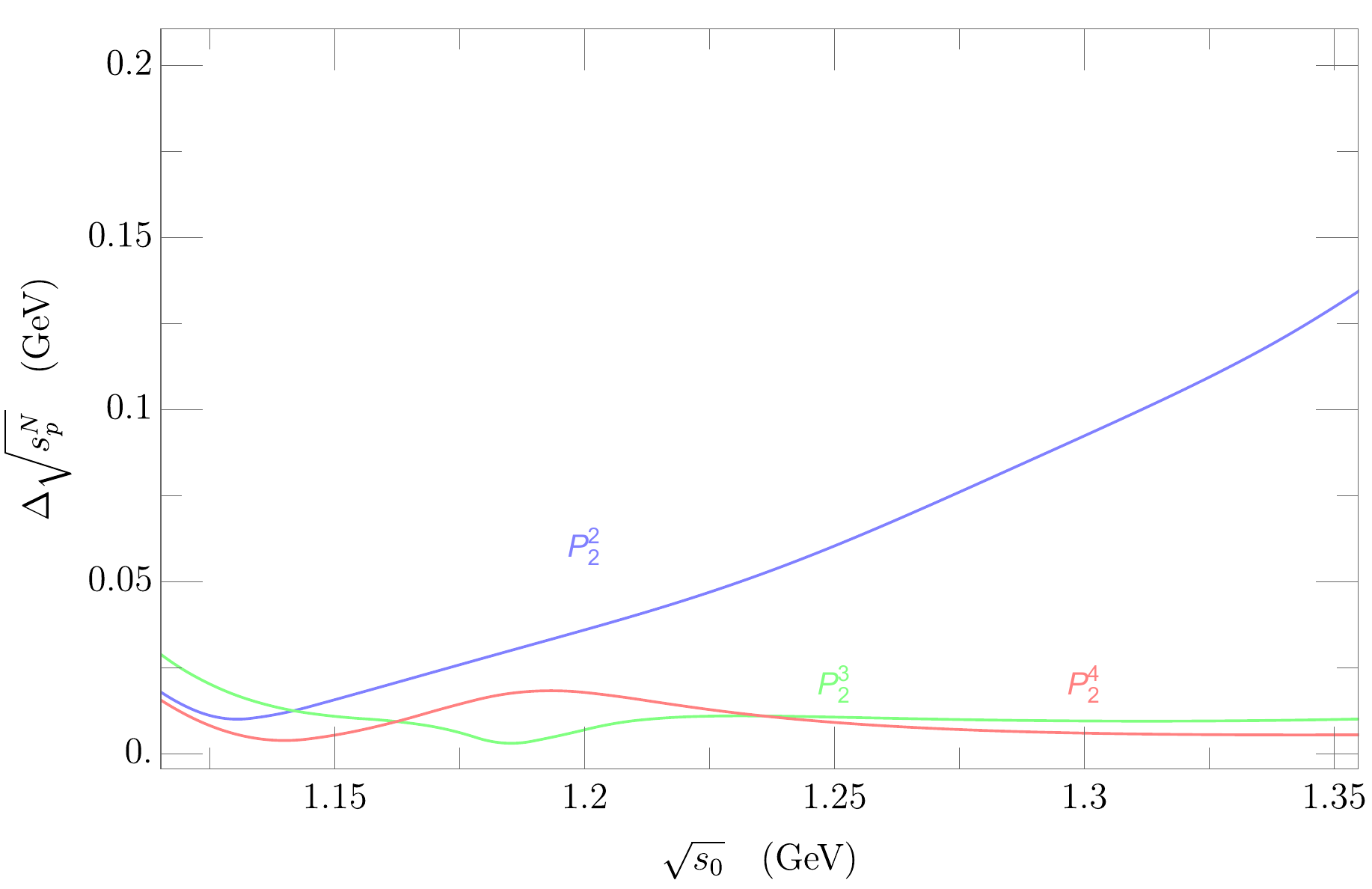}}
\caption{(Top panel) Global parameterizations systematic uncertainties, associated to the truncation of the single pole Pad\'e sequence. (Bottom panel) The same uncertainties for a two-pole Pad\'e sequence.
\label{fig:globalsyst}}
\end{figure} 

\mytitle{Pad\'e Approximants} 
After defining the general $P^N_M$ Pad\'e approximant in the previous subsection,  Eq.~\eqref{eq:Padedef}, we now detail the particular sequences we use, their truncation, as well as how we determine their parameters and uncertainties from the dispersive output or the data parameterizations.

First of all, we choose $s_0$ in~\eqref{eq:Padedef} to center the Taylor expansion defined in Eq.~\eqref{eq:PadeTaylor} in a point of the real axis near the resonance.
The choice of Pad\'e sequence, i.e., the $M$ in $P^N_M$, depends on the analytic structure of the function to be continued. For example, when studying a narrow, isolated resonance, setting $s_0$ near its mass ensures that there is a domain around $s_0$ where  $F(s)$ is analytic but for the pole associated to that resonance. We can then set $M=1$ and thus approximate the amplitude by
\begin{equation}
P_{1}^{N}\left(s, s_{0}\right)=\sum_{k=0}^{N-1} a_{k}\left(s-s_{0}\right)^{k}+\frac{a_{N}\left(s-s_{0}\right)^{N}}{1-\frac{a_{N+1}}{a_{N}}\left(s-s_{0}\right)},
\end{equation}
with $a_{n}=\frac{1}{n !} F^{(n)}\left(s_{0}\right)$ given by the derivatives of the function at $s_0$, as explained above. In this very simple case the pole and its residue are given by
\begin{equation}
s_{p}^{N}=s_{0}+\frac{a_{N}}{a_{N+1}},\quad Z^{N}=-\frac{\left(a_{N}\right)^{N+2}}{\left(a_{N+1}\right)^{N+1}}.
\end{equation}
More often resonances lie close to other non-analytic structures like other poles or thresholds and their cuts. In such cases, we would choose a larger $M$, to mimic those additional singularities.  In this work, we
have considered several possibilities ranging from
$M=1$ to $M=3$. These additional poles, depending on the center, may crudely mimic the effect of the $f_0(980)$, the $f_0(1500)$, or any of the $K\bar K$, $\eta\eta$ or $\rho\rho$ thresholds. In Tab.~2 in the main text we have
provided examples of results for $M=1,2$ and when both sequences converge, they give almost identical results.

 Let us now turn to the calculation of uncertainties for a given center $s_0$.  First, we propagate the uncertainties in the amplitude in the real axis through the Pad\'e coefficients to the pole, and this we call statistical uncertainties.
 
Systematic errors stem from truncating the Pad\'e sequence to a given order $N$.  Naively considering higher orders may approximate the amplitude better, but these require higher derivatives, which propagate larger statistical uncertainties. Given a fixed center $s_0$ and the poles extracted at order $N$ and $N-1$, we will estimate the systematic uncertainty associated with the truncation of the series as
\begin{equation}
\Delta \sqrt{s_{p}^{N}}=\left|\sqrt{s_{p}^{N}}-\sqrt{s_{p}^{N-1}}\right|,
\end{equation}
which is shown for different centers in Fig.~\ref{fig:globalsyst} using as input the ``Global 1" parameterization. In the upper panel, we show that, for the $P^N_1$ sequence, this uncertainty decreases as $N$ increases and that there is a plateau for the optimal choice of center around 1.2-1.25 GeV, reaching the $\sim 10\,$MeV level. In the lower panel, we show a similar plot for the $P^N_2$ sequence, which seems to converge even faster and over a wider region.
In view of these plots, we then decide to truncate the sequence when this systematic uncertainty becomes smaller than the statistical one.
Then we choose our final $s_0$ as the one that minimizes the sum in quadrature of the statistical and systematic uncertainties.

Finally, there is an additional subtlety on how to determine the coefficients of each Pad\'e approximant. 

    \mytitle { Pad\'e approximants from derivatives} These should be used only when we have an analytic formula for the parameterization and its derivatives, like in the ``Global'' case. This approach has been extensively used in the recent past with great success~\cite{Masjuan:2013jha,Masjuan:2014psa,Caprini:2016uxy,Pelaez:2016klv}. Let us briefly summarize the main details here. As can be seen in Fig.~\ref{fig:globalsyst} the $P^N_M$ Pad\'e sequence for one or two poles seems to converge much better to the $f_0(1370)$ pole when $N>3$ and 2, respectively. The preferred center of the expansion lies around $1.2-1.25$ GeV for single-pole approximants, and above $1.35$ GeV for two-pole ones. Notice that already for $P^4_1$ and $P^3_2$ the systematic errors are small compared to the total error listed in Tab.~1 in the main text. When using a two-pole Pad\'e approximant we always find a second pole in the area. 

\begin{figure*}
{\centering
 \raisebox{-0.5\height}{\includegraphics[width=0.45\textwidth]{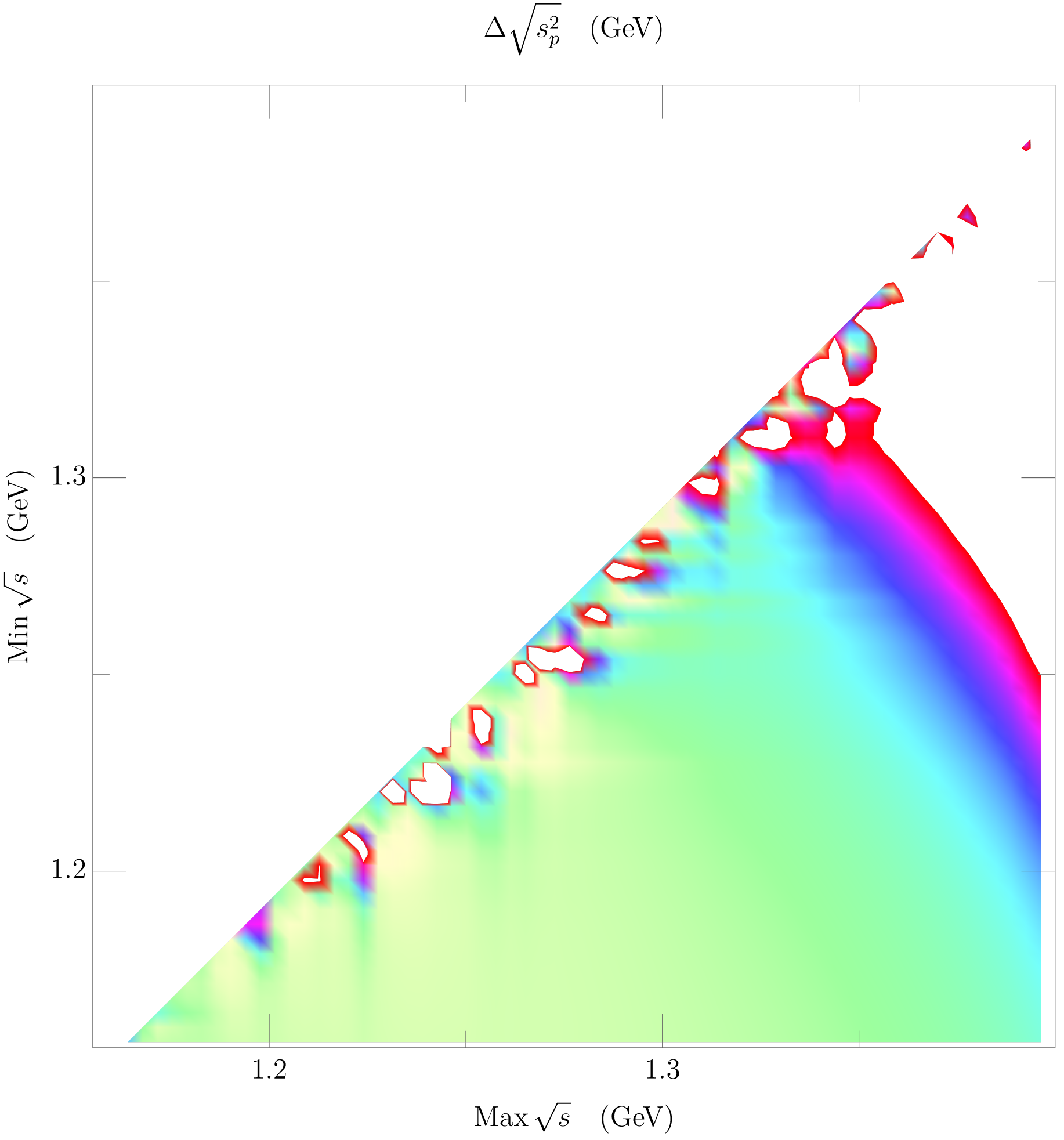} \includegraphics[width=0.45\textwidth]{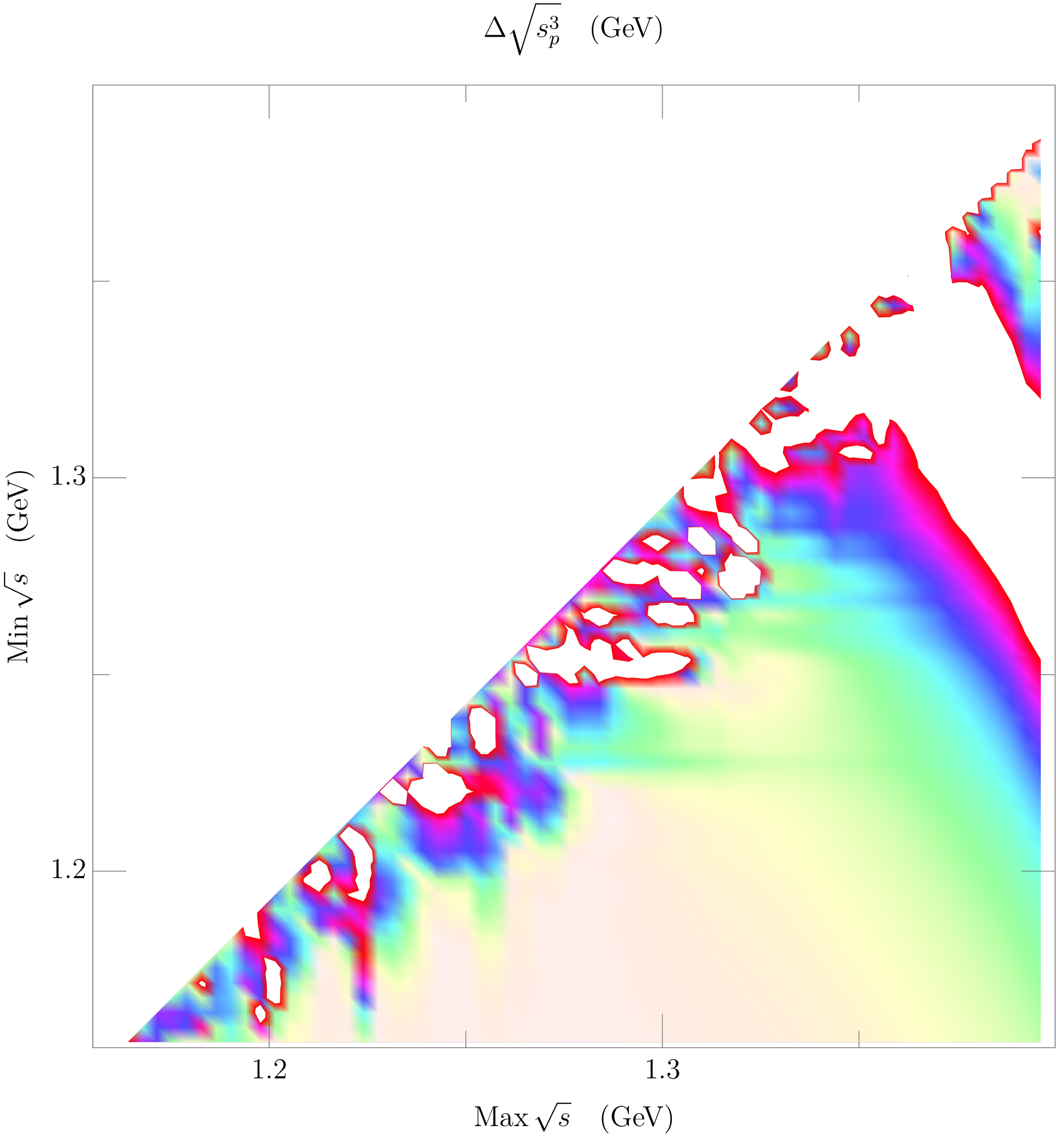}}  \raisebox{-0.5\height}{\includegraphics[width=0.04\textwidth]{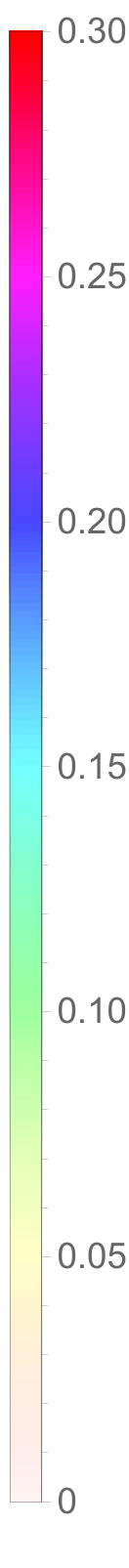}}}
\caption{GKPY+Global Pad\'e fit systematic uncertainties. We show the $P^2_1$ (left) and $P^3_1$ (right) systematic uncertainties associated with the selection of the minimum and maximum values of $\sqrt{s}$ fitted. Notice how the $P^3_1$ fit produces a much smaller systematic uncertainty for a broad region.
\label{fig:royglobalsyst}}
\end{figure*}

    \mytitle { Pad\'e approximants fitted to the amplitudes} Another possible approach is to perform a Pad\'e fit to the amplitude of interest~\cite{Ropertz:2018stk}. This is well suited when we do not know the analytic formula for the derivatives, like when using dispersion relations for the ``GKPY+CFD'' or ``GKPY+Global'' results. In this particular case, the systematic spread is different. We no longer care about the center of the expansion $s_0$, but about the initial and final energy values to be fit. Thus, we will study a vast energy region and select the interval where our Pad\'e sequence converges faster. Once again, we will consider the systematic uncertainty as the difference between the pole positions calculated at two consecutive orders, for the same fitted segment. For illustration, we show in Fig.~\ref{fig:royglobalsyst} the systematic uncertainties for $P^2_1,\, P^3_1$ when analytically continuing the GKPY equations output,  with the Global1 parameterization as input. Two main features can be noticed, the first one is that $P^3_1$ produces on average a much smaller systematic uncertainty than $P^2_1$, and has a vast region for which it becomes negligible. The second is that this region includes segments of substantial length. Similar behavior is found when using the CFD parameterization. Both seem to favor segments with a length of roughly 0.15-0.2 GeV. Concerning the CFD parameterization, let us recall that it is piecewise in this region~\cite{GarciaMartin:2011cn}, and as such there will always be additional non-analytic structures nearby. Finally note that different sequences and methods produce different pole positions, although as seen in Table~\ref{tab:checks} in the main text,  they are compatible and overlap within uncertainties. This is a result of the sizable parameterization dependence that one always suffers when fitting data.

    Unfortunately, this method does not produce a stable $f_0(1370)$ when using forward dispersion relations. The main reason is that these amplitude dispersion relations include all partial waves with a given isospin combination, not only the scalar one. As a result, the $F^{00}$ includes and is actually dominated in this region by the $f_2(1270)$, which is narrower than our desired $f_0(1370)$. If one is to use a single-pole Pad\'e, it will always find the $f_2(1270)$ as the dominant one. Including more poles in the denominator produces also a broader signal behind the tensor resonance. However, the spread of results is very large. We have found that using continued fractions, which in principle are not restricted regarding the number of poles they produce, is more suitable for this particular case.

\subsection{Forward Dispersion Relations for different isospin combinations}\label{app:fdr}

In the main text, we have stated that the most precise Forward Dispersion Relation (FDR) to extract the isospin zero component is the one for the amplitude
\begin{equation} 
F^{00}(s,0)\equiv(F^0(s,0)+2F^2(s,0))/3,
\end {equation}
where $F^I(s,t)$ are the $\pi\pi$ scattering amplitudes with definite isospin $I$. 
Its small uncertainties are due to the positivity of all integrand contributions \cite{GarciaMartin:2011cn,NavarroPerez:2015gaz}. 
Other FDRs were also considered in these references, particularly the $F^{0+}=(F^1+F^2)/3$, also with good positivity properties, and the $F^{I_t=1}$, which has the advantage of being dominated by the $\rho$ reggeon exchange at high energies, without the Pomeron contribution that dominates the other two.

\begin{figure}
\includegraphics[width=0.45\textwidth]{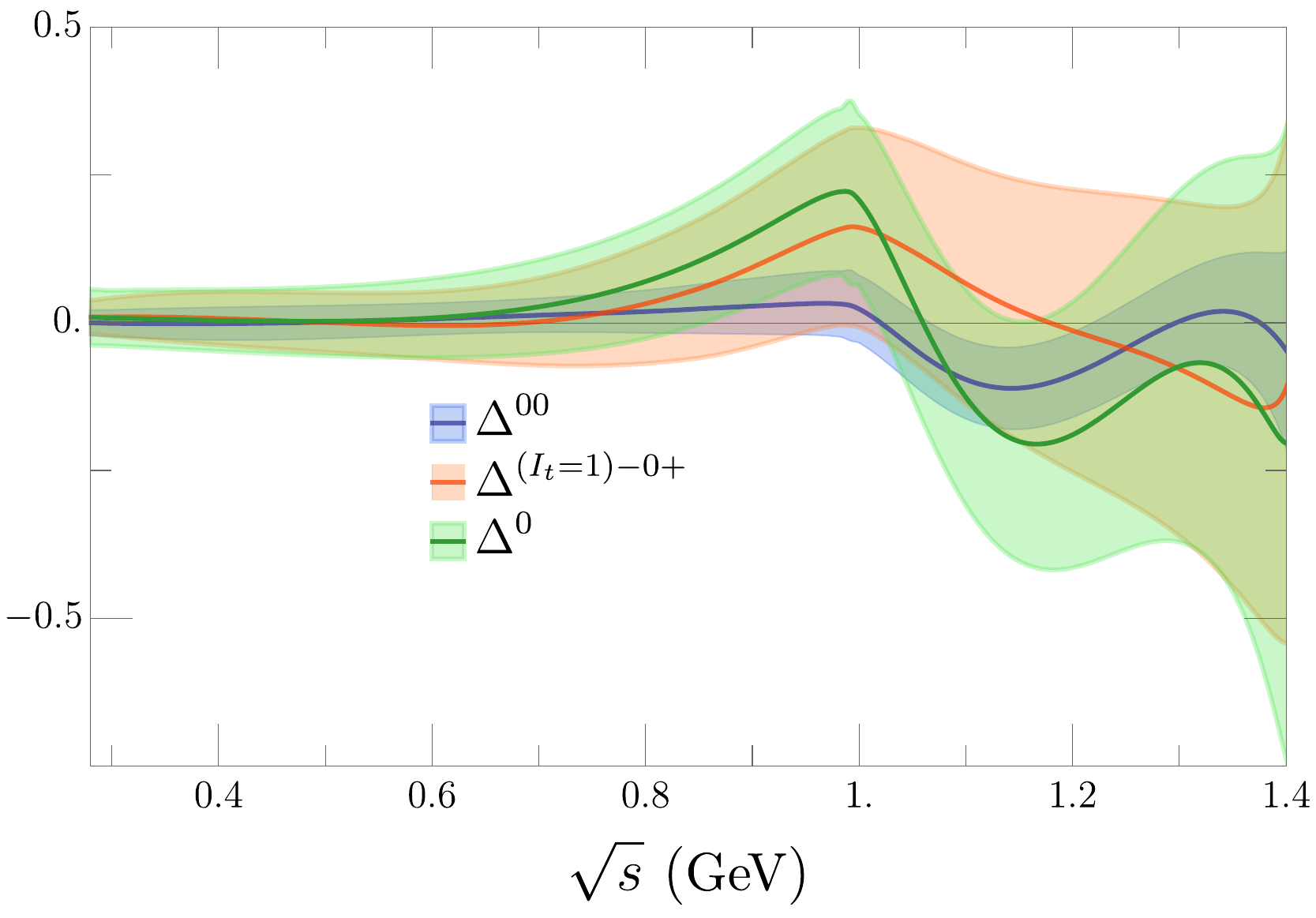} 
\caption{ Differences between the Input and FDR output for three different isospin combinations. The blue line corresponds to the difference for the $F^{00}$ amplitude, used in this work. The orange and green lines correspond to the $F^{I_t=1}-F^{0+}$ and $F^0$ amplitudes respectively. The colored bands represent the relative uncertainties between Input and FDR output.  }\label{Fig:diffFDR}
\end{figure}

Then, in Fig.~\ref{Fig:diffFDR}, we show the difference between input and output for several FDRs and their uncertainty bands. 
In particular, one could consider extracting the pure $F^0$ component. In terms of the $F^{00}$, $F^{0+}$ and $F^{I_t=1}$ FDRs provided in  \cite{GarciaMartin:2011cn,NavarroPerez:2015gaz}, this corresponds to $F^0=2F^{00}-F^{0+}+F^{I_t=1}$. We thus lose positivity in the integrands and the resulting uncertainty, shown in green, becomes larger than for $F^{00}$. The $F^{0+}$ does not contain isospin 0, but can be used to remove the $F^1$ contribution from $F^{I_t=1}$, thus avoiding the presence of yet another resonance pole in the region of interest (the $\rho(1450)$). The fulfillment of this FDR is plotted in Fig.~\ref{Fig:diffFDR} as a light red band, once again much larger than for $F^{00}$. Any other combination will thus suffer from the loss of positivity or the presence of the $F^1$ component and therefore a larger uncertainty than $F^{00}$, which is, therefore, the best choice for our calculations.

Note that FDRs have input from data up to several tens of GeV, but above 1.42 GeV they were not imposed as constraints. The reason is that above 1.42 GeV data comes from total cross-section data described with a Regge parameterization \cite{GarciaMartin:2011cn}, which provides $\im F(s,0)$ via the optical theorem. However, given the large, and monotonously growing, FDR uncertainties, they are expected to remain valid within uncertainties well above that energy.

\subsection{On the stability of our analytic continuation}

 \begin{figure*}
{\centering
\includegraphics[width=0.9\textwidth]{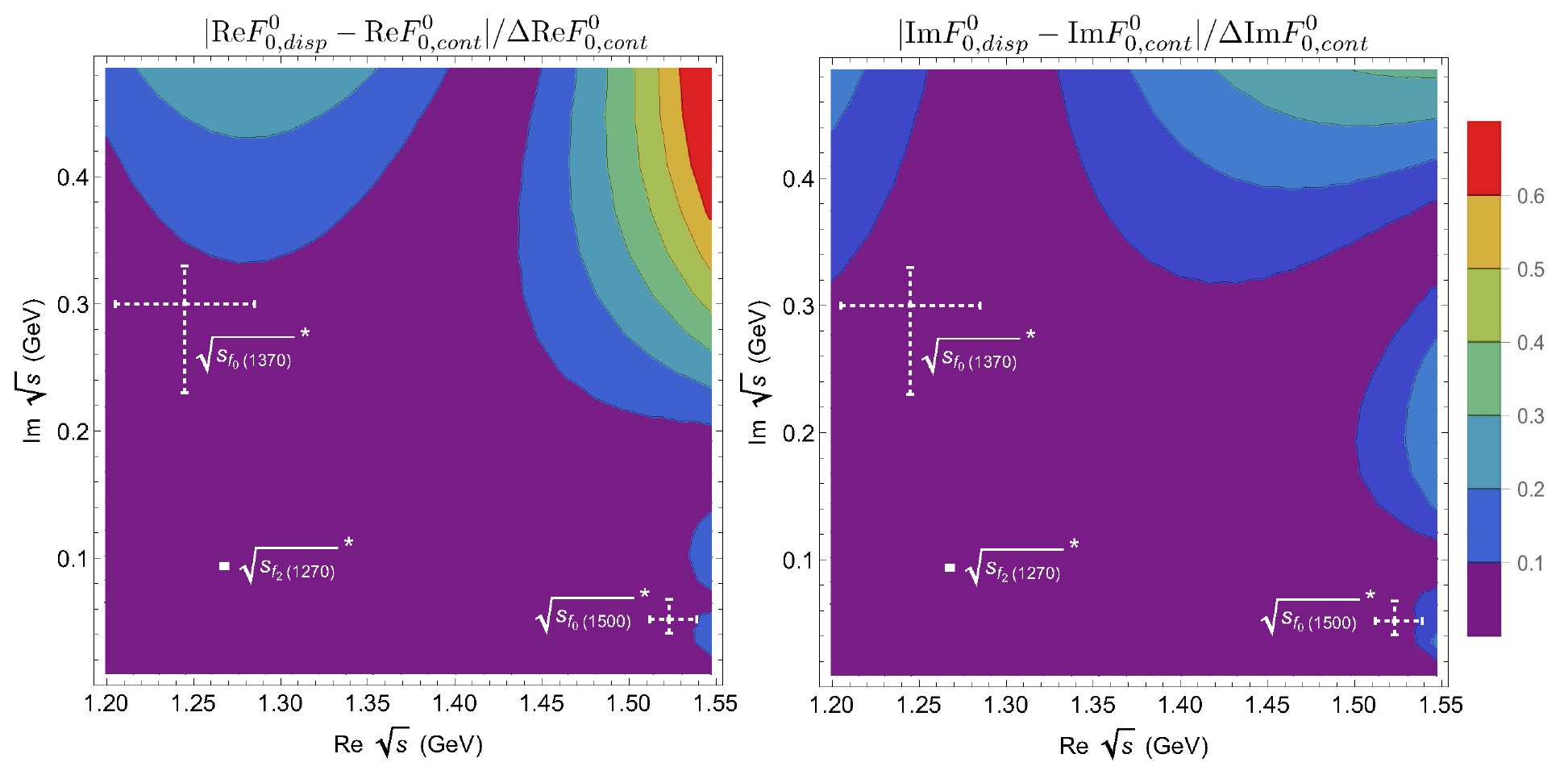}
\caption{Absolute value of the difference between the $F^0_0$ calculated dispersively from its FDR and the continued fraction analytic continuation, divided by the uncertainty on the latter. As explained in the manuscript, the continued fractions are built in the real axis by using a discretized segment with $N$ interpolation points.  We show the upper half complex plane in the first Riemann sheet. We plot the real part on the left panel and the imaginary on the right one. For illustration, we show the conjugate of the pole position for each one of the resonance poles we found in the text, which of course do not lie here but in the proximal sheet.}\label{fig:complex}}
\end{figure*}

As is well known, for two functions that differ very slightly in a segment in the real axis, there is always a value in the complex plane where their analytic continuations may differ by more than any given value. This 
problem for analytic continuations, which has been one of our main concerns, is the reason why we have performed so many different checks. We use several continuation methods and different parameterizations of the input. Let us recall that from these we have estimated our systematic uncertainties. Each method and parameterization produce a stable analytic continuation in the region of interest. Once their pole position has been obtained, statistical uncertainties are produced. 


Let us first discuss our main results obtained via continued fractions. For a given segment on the real axis, we vary the number $N$ of interpolation points of our continued fractions. Note, as stated above, that for a given $N$ there can be as many as $N/2$ poles in $C_N$, but as we show in Fig.~3 in the manuscript, there are only three poles that appear consistently in the region of interest. Their values are very stable under changes of $N$. We plot up to $N\sim50$, including their differences for the pole parameters as uncertainty. Moreover, sampling $N$'s as large as 100, or varying the size of the segment in the real axis produce only slight variations.

As commented in subsection~\ref{app:results}, only a handful of times are some of these poles not detected while resampling. Partly this is because our pole search algorithm fails, but sometimes it is because other spurious poles appear. In such cases, the function interpolated in the real segment is still within errors of the FDR output, but we get a bizarre continuation with untrustworthy poles. However, these spurious poles move erratically with little parameter variation. They are not found consistently throughout the sampling and are thus numerical noise.

 Furthermore, in order to estimate the parameterization bias, we consider four different inputs for the forward dispersion relations, i.e., the piecewise CFD and the analytic ``Global parameterizations". These are different in the real axis, although consistent within uncertainties. Once again they produce only slightly different $f_0(1370)$ poles, becoming part of our systematic uncertainty.

 Additionally, for $\pipi$ scattering, we analyze the GKPY dispersive outputs used in the real axis to compare them with the forward dispersive results. This is a different dispersive output since the GKPY equations provide the value of the $f_0^0$ wave. Since the $f_2(1270)$ is not present in $f_0^0$, one can apply the Padé sequence method instead of continued fractions. As can be seen in Table~\ref{tab:checks} in the main text, they are compatible with our main result and thus produce negligible systematic effects. Even more importantly, the numerical instabilities in isolated samples of parameters with continued fractions are not found here.

 Finally, since the global parameterizations also contain a pole associated with the $f_0(1370)$, one can compare the different pole extractions produced between the analytic results, those obtained from Padé approximants, and those coming from continued fractions. Once again, they produce compatible results, with comparable error estimates. 
 
 The different nature of the methods and parameterizations used to perform the analytic continuation showcase how, even for the $f_0(1370)$, the extraction is robust. Nonetheless, there is still room for one final check. So far we used these methods to look at the proximal sheet in the lower half plane. However, these continuation methods simultaneously describe the first Riemann sheet. This sheet is accurately known for the full amplitudes, as it can be calculated explicitly using the FDRs integral form. In Fig.~\ref{fig:complex} we show the difference between the FDRs calculated rigorously with the dispersion relation and the continued fraction analytic continuation from the segment of reference (1.2-1.4 GeV), computed, all in units of the uncertainty. For the continued fraction method we have followed at each point the same procedure as we have used in the main text to determine the $f_0(1370)$ pole parameters: using a sequence of $C_N$ up to N=50, propagate the statistical uncertainties and add a systematic uncertainty to cover the results of the whole sequence.
 We see that for the most part of the region of interest, the continued fraction result coincides with the dispersive one within $10\%$ of the estimated uncertainty of the continuation method. To reach a difference in the order of the uncertainty we have to go at least beyond 1.5 GeV in the real axis and much deeper in the complex plane than the region we are interested in. Varying the number of interpolating points or the endpoints of the segment produces only slight variations in the outcome.

Thus, we achieved a remarkable agreement between the continued fraction method and the FDRs in the complex plane, as well as the agreement between different extrapolation methods and inputs. This makes our uncertainty estimates the most robust in the literature.

\color{black}

\bibliographystyle{apsrev4-1}
\bibliography{largebiblio.bib}

\newpage

\end{document}